# Damping dependence of spin-torque effects in thermally assisted magnetization reversal

Y.P. Kalmykov,[1] D. Byrne,[2] W.T. Coffey,[3] W. J. Dowling,[3] S.V.Titov,[4] and J.E. Wegrowe[5]

[1]Univ. Perpignan Via Domitia, Laboratoire de Mathématiques et Physique, F-66860, Perpignan, France
[2]School of Physics, University College Dublin, Belfield, Dublin 4, Ireland
[3]Department of Electronic and Electrical Engineering, Trinity College, Dublin 2, Ireland
[4]Kotel'nikov Institute of Radio Engineering and Electronics of the Russian Academy of Sciences, Vvedenskii Square 1, Fryazino, Moscow Region, 141120, Russia
[5]Laboratoire des Solides Irradiés, Ecole Polytechnique, 91128 Palaiseau Cedex, France

**Thermal fluctuations of nanomagnets driven by spin-polarized currents are treated via the Landau-Lifshitz-Gilbert equation as generalized to include both the random thermal noise field and Slonczewski spin-transfer torque (STT) terms. The magnetization reversal time of such a nanomagnet is then evaluated for wide ranges of damping by using a method which generalizes the solution of the so-called Kramers turnover problem for mechanical Brownian particles thereby bridging the very low damping (VLD) and intermediate damping (ID) Kramers escape rates, to the analogous magnetic turnover problem. The reversal time is then evaluated for a nanomagnet with the free energy density given in the standard form of superimposed easy-plane and in-plane easy-axis anisotropies with the dc bias field along the easy axis.**

## I. Introduction

Due to the spin-transfer torque (STT) effect [1-6], the magnetization of a nanoscale ferromagnet *may be altered by spin-polarized currents*. This phenomenon occurs because an electric current with spin polarization in a ferromagnet has an associated flow of angular momentum [3,7] thereby exerting a macroscopic spin torque. The phenomenon is the origin of the novel subject of spintronics [7,8], i.e., current-induced control over magnetic nanostructures. Common applications are very high-speed current-induced magnetization switching by (a) reversing the orientation of magnetic bits [3,9] and (b) using spin polarized currents to control steady state microwave oscillations [9]. This is accomplished via the steady state magnetization precession due to STT representing the conversion of DC input into an AC output voltage [3]. Unfortunately, thermal fluctuations cannot now be ignored due to the *nanometric* size of STT devices, e.g., leading to mainly noise-induced switching at currents far less than the critical switching current without noise [10] as corroborated by experiments (e.g., [11]) demonstrating that STT near room temperature significantly alters thermally activated switching processes. These now exhibit a pronounced dependence on both material and geometrical parameters. Consequently, an accurate account of STT switching effects at finite temperatures is necessary in order to achieve further improvements in the design and interpretation of experiments, in view of the manifold practical applications in spintronics, random access memory technology, and so on.

During the last decade, various analytical and numerical approaches to the study of STT effects in the thermally assisted magnetization reversal (or switching) time in nanoscale ferromagnets have been developed [6,7,12-26]. Their objective being to generalize methods *originally developed for zero STT* [12,27-32] such as stochastic dynamics simulations (e.g., Refs. [21-25]) and extensions to spin Hamiltonians of the mean first passage time (MFPT) method (e.g., Refs. [16] and [17]) in the Kramers escape rate theory [33,34]. However, unlike zero STT substantial progress in escape rate theory including STT effects has so far been achieved only in the limit of very low damping (VLD), corresponding to vanishingly small values of the damping parameter $\alpha$ in the Landau-Lifshitz-Gilbert-Slonczewski equation (see Eq. (5) below). Here the pronounced time separation between *fast* precessional and *slow* energy changes in *lightly* damped closed phase space trajectories (called Stoner-Wohlfarth orbits) has been exploited in Refs. [7,14,16,17] to formulate a one-dimensional Fokker-Planck equation for the energy distribution function which may be solved by quadratures. This equation is essentially similar to that derived by Kramers [33] in treating the VLD noise-activated escape rate of a point Brownian particle from a potential well although the Hamiltonian of the magnetic problem is no longer separable and additive and the barrier height is now STT dependent. The Stoner-Wohlfarth orbits and steady precession along such an orbit of constant energy occur if the spin-torque is strong enough to cancel out the dissipative torque. The origin of the orbits arises from the bistable (or, indeed, in general multistable) structure of the anisotropy potential. This structure allows one to define a nonconservative "effective" potential with damping- and current-dependent potential barriers between stationary self-oscillatory states of the magnetization, thereby permitting one to estimate the reversal (switching) time between these states. The magnetization reversal time in the VLD limit is then evaluated [16,17,35] both for *zero and nonzero STT.* In particular, for *nonzero STT*, the VLD reversal time has been evaluated analytically in Refs. [16,17]. Here it has been shown that in the high barrier limit, an asymptotic equation for the VLD magnetization reversal time from a *single* well in the presence of the STT is given by

$$\tau^{\mathrm{VLD}} \approx \frac{1}{\alpha S_{E_C} \Gamma^{\mathrm{TST}}}. \tag{1}$$

In Eq. (1), $\alpha$ is the damping parameter arising from the surroundings, $\Gamma^{\mathrm{TST}} = f_{E_A} e^{-\Delta E}$ is the escape rate rendered by transition state theory (TST) which ignores effects due to the loss of spins at the barrier [34], $f_{E_A}$ is the well precession frequency, $\Delta E$ is the damping and spin-polarized-current dependent effective energy barrier, and $S_{E_C}$ is the dimensionless action at the saddle point *C* (the action is given by Eq. (13) below).

The most essential feature of the results obtained in Refs. [16,17,35] and how they pertain to this paper is that they apply at VLD only where the inequality $\alpha S_{E_C} << 1$ holds meaning that the energy loss per cycle of the almost periodic motion at the critical energy is much less than the thermal energy. Unfortunately for typical values of the material parameters $S_{E_C}$ may be very high ($>10^3$), meaning that this inequality can be fulfilled only for $\alpha < 0.001$. In addition, both experimental and theoretical estimates suggest higher values of $\alpha$ of the order of 0.001-0.1 (see, e.g., Refs. [6,36-38]), implying that the VLD asymptotic results are no longer valid as they will now differ substantially from the true value of the reversal time. These considerations suggest that the asymptotic calculations for STT should be extended to include both the VLD and intermediate damping (ID) regions. This is our primary objective here. Now like point Brownian particles which are governed by a separable and additive Hamiltonian, in the escape rate problem as it pertains to magnetic moments of nanoparticles, *three* regimes of damping appear [12,33,34]. These are (i) very low damping $(\alpha S_{E_C} << 1)$, (ii) intermediate-to-high damping (IHD) $(\alpha S_{E_C} >> 1)$, and (iii) a more or less critically damped turnover regime $(\alpha S_{E_C} \sim 1)$. Also, Kramers [33] obtained his now-famous VLD and IHD escape rate formulas for point Brownian particles by assuming in both cases that the energy barrier is much greater than the thermal energy so that the concept of an escape rate applies. He mentioned, however, that he could not find a general method of attack in order to obtain an escape rate formula valid for any damping regime. This problem, namely the Kramers turnover, was initially solved by Mel'nikov and Meshkov [39]. They obtained an escape rate that is valid for all values of the damping by a semi heuristic argument, thus constituting a solution of the Kramers turnover problem for point particles. Later, Grabert [40] and Pollak *et al.* [41] have presented by using a coupled oscillator model of the thermal bath, a complete solution of the Kramers turnover problem and have shown that the turnover escape rate formula can be obtained without the *ad hoc* interpolation between the VLD and IHD regimes as used by Mel'nikov and Meshkov. Finally, Coffey *et al.* [42,43] have shown for classical spins that *at zero STT*, the magnetization reversal time for values of damping up to intermediate values, $\alpha \le 1$, can also be evaluated via the turnover formula for the escape rate bridging the VLD and ID escape rates, namely,

$$\tau \approx \frac{1}{\Gamma^{\mathrm{TST}} A(\alpha S_{E_C})}, \tag{2}$$

where $A(z)$ is the so-called depopulation factor, namely [39-42]

$$A(z) = e^{\frac{1}{\pi}\int_0^\infty \frac{\ln\left\{1-\exp[-z(\lambda^2+1/4)]\right\}}{\lambda^2+1/4} d\lambda}. \tag{3}$$

Now the ID reversal time (or the lower bound of the reversal time) may always be evaluated via TST as [32,34]

$$\tau^{\mathrm{ID}} \approx \frac{1}{\Gamma^{\mathrm{TST}}}. \tag{4}$$

Therefore because $A(\alpha S_{E_C}) \to \alpha S_{E_C}$ is the energy loss per cycle at the critical energy $\alpha S_{E_C} \to 0$ [39] (i.e., in the VLD limit), Eq. (2) transparently reduces to the VLD Kramers result, Eq. (1). Moreover in the ID range, where $A(\alpha S_{E_C}) \approx 1$, Eq. (2) reduces to the TST Eq. (4). Nevertheless in the high barrier limit $S_{E_C} >> 1$, $\tau$ given by Eq. (2) can *substantially* deviate in the damping range $0.001 < \alpha < 1$ both from $\tau^{\mathrm{ID}}$, Eq. (4), and $\tau^{\mathrm{VLD}}$, Eq. (1). Now, the approach of Coffey *et al.* [42,43] generalizing the Kramers turnover results to classical spins (nanomagnets) was developed for zero STT, nevertheless, it can also be used to account for STT effects. Here we shall extend the zero STT results of Refs. [14,16,17,39-42] treating the *damping dependence of STT effects* in the magnetization reversal of nanoscaled ferromagnets via escape rate theory in the most important range of damping comprising the VLD and ID ranges, $\alpha \le 1$.

## II. Model

The object of our study is the role played by STT effects in the thermally assisted magnetization reversal using an adaptation of the theory of thermal fluctuations in nanomagnets developed in the seminal works of Néel [27] and Brown [28,29]. The Néel-Brown theory is effectively an adaptation of the Kramers theory [33,34] originally given for point Brownian particles to magnetization relaxation governed by a gyromagnetic-like equation which is taken as the Langevin equation of the process. Hence, the verification of that theory in the pure (i.e., without STT) nanomagnet context nicely illustrates the Kramers conception of a thermal relaxation process as escape over a potential barrier arising from the shuttling action of the Brownian motion. However, it should be recalled throughout that unlike nanomagnets at zero STT (where the giant spin escape rate theory may be effectively regarded as fully developed), devices based on STT, due to the injection of the spin-polarized current, invariably represent an *open* system in an *out-of-equilibrium steady state.* This is in marked contrast to the conventional steady state of nanostructures characterized by the Boltzmann equilibrium distribution that arises when STT is omitted. Hence both the governing Fokker-Planck and Langevin equations and the escape rate theory based on these must be modified.

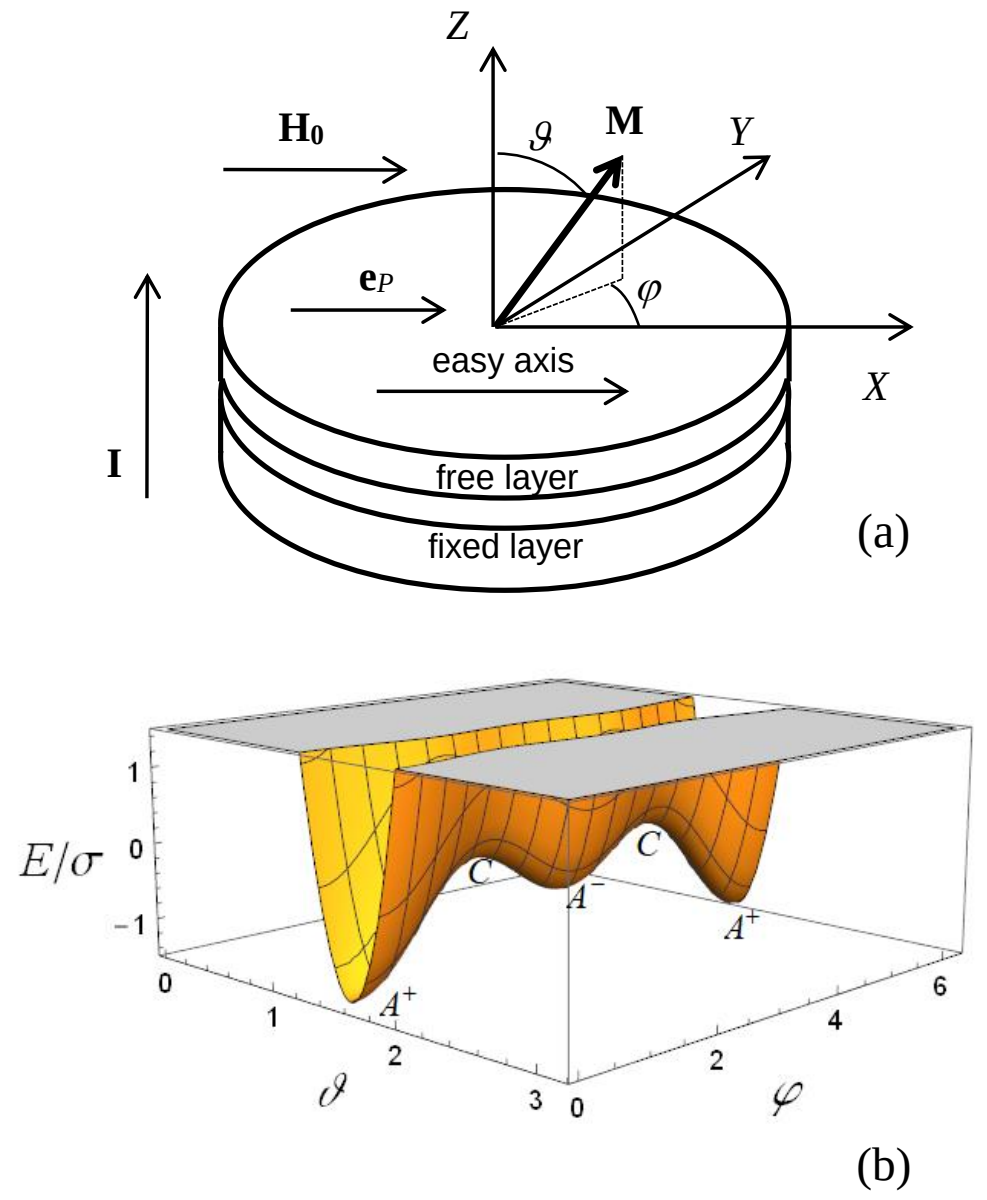

Fig. 1. (a) Geometry of the problem: A STT device consists of two ferromagnetic strata labelled the *free* and *fixed* layers, respectively, and a normal conducting spacer all sandwiched on a pillar between two ohmic contacts [3,6]. Here **I** is the spin-polarized current, **M** is the magnetization of the free layer, $\mathbf{H}_0$ is the dc bias magnetic field. The magnetization of the fixed layer is directed along the unit vector $\mathbf{e}_P$. (b) Free energy potential of the free layer presented in the standard form of superimposed easy-plane and in-plane easy-axis anisotropies, Eq. (7), at $\delta = 20$ and $h = 0.2$.

To facilitate our discussion, we first describe a schematic model of the STT effect. The archetypal model (Fig. 1 (a)) of a STT device is a nanostructure comprising two magnetic strata labeled the *free* and *fixed* layers and a nonmagnetic conducting spacer. The fixed layer is much more strongly pinned along its orientation than the free one. If an electric current is passsed through the fixed layer it becomes spin-polarized. Thus, the current, as it encounters the free layer, induces a STT. Hence, the magnetization $\mathbf{M}$ of the free layer is altered. Both ferromagnetic layers are assumed to be *uniformly* magnetized [3,6]. Although this giant coherent spin approximation cannot explain all observations of the magnetization dynamics in spin-torque systems, nevertheless many qualitative features needed to interpret experimental data are satisfactorily reproduced. Indeed, the current-induced magnetization dynamics in the free layer may be described by the Landau-Lifshitz-Gilbert-Slonczewski equation including thermal fluctuations, i.e., the usual Landau-Lifshitz-Gilbert equation [44] including STT, however augmented by a random magnetic field $\boldsymbol{\eta}(t)$ which is regarded as white noise. Hence it now becomes a magnetic Langevin equation [3,6,7,12], viz.,

$$\dot{\mathbf{u}} = -\gamma\left[\mathbf{u}\times\left(\mathbf{H}+\boldsymbol{\eta}\right)\right] + \alpha\left[\mathbf{u}\times\dot{\mathbf{u}}\right] + \gamma\left[\mathbf{u}\times\left[\mathbf{u}\times\mathbf{I}_S\right]\right]. \quad (5)$$

Here $\mathbf{u} = \mathbf{M}/M_S$ is the unit vector directed along $\mathbf{M}$, $M_S$ is the saturation magnetization, and $\gamma$ is the gyromagnetic-type constant. The *effective* magnetic field $\mathbf{H}$ comprising the anisotropy and external applied fields is defined as

$$\mathbf{H} = -\frac{kT}{v\mu_0 M_S}\frac{\partial E}{\partial \mathbf{u}}. \quad (6)$$

Here $E$ is the normalized free energy density of the free layer constituting a conservative potential, $v$ is the free layer volume, $\mu_0 = 4\pi\cdot 10^{-7}\,\mathrm{JA^{-2}m^{-1}}$ in SI units, and $kT$ is the thermal energy. For purposes of illustration, we shall take $E(\vartheta,\varphi)$ in the standard form of superimposed easy-plane and in-plane easy-axis anisotropies plus the Zeeman term due to the applied magnetic field $\mathbf{H}_0$ [45] (in our notation):

$$E(\vartheta,\varphi) = -\sigma(\sin^2\vartheta\cos^2\varphi - \delta\cos^2\vartheta + 2h\sin\vartheta\cos\varphi). \quad (7)$$

In Eq. (7) $\vartheta$ and $\varphi$ are the polar and azimuthal angles in the usual spherical polar coordinate system, $h = H_0/(2M_S D_\parallel)$ and $\sigma = v\mu_0 M_S^2 D_\parallel/(kT)$ are the external field and anisotropy parameters, $\delta = D_\perp / D_\parallel >> 1$ is the biaxiality parameter characterized by $D_\parallel$ and $D_\perp$ thereby encompassing both demagnetizing and magnetocrystalline anisotropy effects (since $\sigma$ and $\delta$ are determined by both the volume and the thickness of the free layer, their numerical values may vary through a very large range, in particular, they can be very large, > 100 [45]). The form of Eq. (7) implies that both the applied field $\mathbf{H}_0$ and the unit vector $\mathbf{e}_P$ identifying the magnetization direction in the fixed layer are directed along the easy $X$-axis (see Fig. 1(a)). In general, $E(\vartheta,\varphi)$ as rendered by Eq. (7) has two *equivalent saddle points C* and two *nonequivalent wells* at $A^+$ and $A^-$ (see Fig.1(b)). Finally, the STT induced field $\mathbf{I}_S$ is given by

$$\mathbf{I}_S = \frac{kT}{v\mu_0 M_S}\frac{\partial \Phi}{\partial \mathbf{u}}, \quad (8)$$

where $\Phi$ is the normalized nonconservative potential due to the spin-polarized current, which in its simplest form is

$$\Phi(\vartheta,\varphi) = J\left(\mathbf{e}_P\cdot\mathbf{u}\right). \quad (9)$$

In Eq. (9), $J = b_P \hbar I/(|e|kT)$ is the dimensionless STT parameter, $I$ is the spin-polarized current regarded as positive if electrons flow from the free into the fixed layer, $e$ is the electronic charge, $\hbar$ is Planck's reduced constant, and $b_P$ is a parameter determined by the spin polarization factor $P$ [1]. Accompanying the magnetic Langevin equation (5) (i.e., the stochastic differential equation of the random magnetization process), one has the Fokker-Planck equation for the evolution of the associated probability density function $W(\vartheta,\varphi,t)$ of orientations of $\mathbf{M}$ on the unit sphere, viz., [6,12,16]

$$\frac{\partial W}{\partial t} = \mathrm{L}_{\mathrm{FP}} W \,, \tag{10}$$

where $\mathrm{L}_{\mathrm{FP}}$ is the Fokker-Planck operator in phase space $(\vartheta, \varphi)$ defined via [6,12,26]

$$\mathrm{L}_{\mathrm{FP}} W = \frac{1}{2\tau_{\mathrm{N}} \sin\vartheta} \left\{ \frac{\partial}{\partial\vartheta} \sin\vartheta \left[ \frac{\partial W}{\partial\vartheta} + W \left( \frac{\partial (E + \alpha^{-1}\Phi)}{\partial\vartheta} + \frac{1}{\sin\vartheta} \frac{\partial(\alpha^{-1}E - \Phi)}{\partial\varphi} \right) \right] + \frac{1}{\sin\vartheta} \frac{\partial}{\partial\varphi} \left[ \frac{\partial W}{\partial\varphi} + W \left( \frac{\partial (E + \alpha^{-1}\Phi)}{\partial\varphi} - \sin\vartheta \frac{\partial(\alpha^{-1}E - \Phi)}{\partial\vartheta} \right) \right] \right\} \tag{11}$$

and $\tau_{\mathrm{N}} = v\mu_0 M_{\mathrm{S}} (\alpha + \alpha^{-1}) / (2\gamma kT)$ is the free diffusion time of the magnetic moment. If $\Phi = 0$ (zero STT), Eq. (10) becomes the original Fokker-Planck equation derived by Brown [33] for magnetic nanoparticles.

## III. ESCAPE RATES AND REVERSAL TIME IN THE DAMPING RANGE $\alpha \le 1$

The magnetization reversal time can be calculated exactly by evaluating the smallest nonvanishing eigenvalue $\lambda_1$ of the Fokker-Planck operator $\mathrm{L}_{\mathrm{FP}}$ in Eq. (10) [32,34,42]. Thus $\lambda_1$ is the inverse of the longest relaxation time of the magnetization $\tau = 1/\lambda_1$, which is usually associated with the reversal time. In the manner of zero STT [42,43], the calculation of $\lambda_1$ can be approximately accomplished using the Mel'nikov-Meshkov formalism [39]. This relies on the fact that in the *high barrier* and *underdamped* limits, one may rewrite the Fokker-Planck equation, Eq. (10), as an energy-action diffusion equation. This in turn is very similar to that for translating point Brownian particles moving along the *x*-axis in an external potential $V(x)$ [7,17,42]. In the underdamped case, which is the range of interest, for the escape of spins from a *single* potential well with a minimum at a point *A* of the magnetocristalline anisotropy over a single saddle point *C*, the energy distribution function $W(E)$ for magnetic moments precessing in the potential well can then be found via an integral equation [42], which can be solved for $W(E)$ by the Wiener–Hopf method. Then, the flux-over-population method [33,34] yields the decay (escape) rate as $\tau^{-1} = J_C / N_A$. Here $J_C = \mathrm{const}$ is the probability current density over the saddle point and $N_A = \int_{E_A}^{E_C} W(E) dE$ is the well population while the escape rate is rendered as the product of the depopulation factor $A(\alpha S_{E_C})$, Eq. (3), and the TST escape rate $\Gamma^{\mathrm{TST}} = f_{E_A} e^{-\Delta E}$. In the preceding equation $\Delta E$ is the effective spin-polarized current dependent energy barrier given by

$$\Delta E = E_C - E_A - \frac{1}{\alpha} \int_{E_A}^{E_C} \frac{V_E}{S_E} dE \,, \tag{12}$$

where $E_A$ is the energy at the bottom of the potential well, $E_C$ is the energy at the saddle point, and the dimensionless action $S_E$ and the dimensionless work $V_E$ done by the STT are defined as [7,17]

$$S_E = \oint_E \left( \left[ \frac{\partial E}{\partial \mathbf{u}} \times \mathbf{u} \right] \cdot d\mathbf{u} \right), \tag{13}$$

$$V_E = \oint_E \left( \left[ \frac{\partial \Phi}{\partial \mathbf{u}} \times \mathbf{u} \right] \cdot d\mathbf{u} \right) \,, \tag{14}$$

respectively. The contour integrals in Eqs. (13) and (14) are taken along the energy trajectory $E = \mathrm{const}$ and are to be evaluated in the vanishing damping sense.

For the bistable potential, Eq. (7), having two *nonequivalent wells* $A^+$ and $A^-$ with minima $E^{\pm} = \sigma(-1 \mp 2h)$ at $\varphi_{A^+} = 0$ and $\varphi_{A^-} = \pi$, respectively, and two *equivalent saddle points C with* $E_C = \sigma h^2$ at $\cos\varphi_C = -h$ (see Fig. 1(b)) we see that two wells and two escape routes over two saddle points are involved in the relaxation process. Thus, a finite probability for the magnetic dipole to return to the initial well having already visited the second one exists. This possibility cannot be ignored *in the underdamped regime* because then the magnetic dipole having entered the second well loses its energy so slowly that even after several precessions, thermal fluctuations may still reverse it back over the potential barrier. In such a situation, on applying the Mel'nikov-Meshkov formalism [39] to the free energy potential, Eq. (7), and the nonconservative potential, Eq. (9), the energy distribution functions $W_+(E)$ and $W_-(E)$ for magnetic moments precessing in the two potential wells can then be found by solving two coupled integral equations for $W_+(E)$ and $W_-(E)$. These then yield the depopulation factor $A(\alpha S_{E_C}^+, \alpha S_{E_C}^-)$ via the Mel'nikov-Meshkov formula for two wells, viz., [39]

$$A(\alpha S_{E_C}^+, \alpha S_{E_C}^-) = \frac{A(\alpha S_{E_C}^+) A(\alpha S_{E_C}^-)}{A(\alpha S_{E_C}^+ + \alpha S_{E_C}^-)} \,.$$

Here $A(z)$ is the depopulation factor for a single well introduced in accordance with Eq. (3) above while $S_{E_C}^{\pm}$ are the dimensionless actions at the energy saddle points for two wells. These are to be calculated via Eq. (13) by integrating along the energy trajectories $E = E_C^{\pm}$ between two saddle points and are explicitly given by

$$S_{E_C}^{\pm} = \oint_{E_C} \left( \left[ \frac{\partial E}{\partial \mathbf{u}} \times \mathbf{u} \right] \cdot d\mathbf{u} \right) = \frac{4\delta\sigma(1 + \delta - h^2)}{(1+\delta)^{3/2}} \times \left\{ \sqrt{(1-h^2)(1+\delta^{-1})} \pm 2h \arctan \sqrt{\frac{\sqrt{\delta^{-1}(1-h^2)+1} \pm h}{\sqrt{\delta^{-1}(1-h^2)+1} \mp h}} \right\} \tag{15}$$

(at zero dc bias field, $h$ = 0, these simplify to $S_{E_C}^+ = S_{E_C}^- = 4\sqrt{\delta}\sigma$). Furthermore, the overall TST escape rate $\Gamma^{\mathrm{TST}}$ for a bistable potential, Eq.(7), is estimated via the *individual* escape rates $\Gamma_{\pm}^{\mathrm{TST}}$ from each of the two wells as

$$\Gamma^{\mathrm{TST}} = \Gamma_+^{\mathrm{TST}} + \Gamma_-^{\mathrm{TST}} = 2\left( f^+ e^{-\Delta E^+} + f^- e^{-\Delta E^-} \right). \tag{16}$$

In Eq. (16), the factor 2 occurs because two magnetization escape routes from each well over the two saddle points exist, while $\Delta E^{\pm}$ are the effective spin-polarized current dependent barrier heights for two wells (explicit equations for $\Delta E^{\pm}$ are derived in Appendix A). In addition

$$f^{\pm}=\frac{1}{2\pi\tau_0}\sqrt{(1\pm h+\delta)(1\pm h)} \tag{17}$$

are the corresponding well precession frequencies, where $\tau_0=\left(2\gamma M_S D_{\|}\right)^{-1}$ is a precession time constant. Thus, the decay rate $\tau^{-1}$ becomes

$$\tau^{-1}\approx\frac{A(\alpha S_{E_C}^{+})A(\alpha S_{E_C}^{-})}{\pi\tau_0 A(\alpha S_{E_C}^{+}+\alpha S_{E_C}^{-})}\left[\sqrt{(1+h+\delta)(1+h)}e^{-\sigma(1+h)^2+\frac{J}{\alpha}F^{+}(\delta,h)}+\sqrt{(1-h+\delta)(1-h)}e^{-\sigma(1-h)^2-\frac{J}{\alpha}F^{-}(\delta,h)}\right], \tag{18}$$

where both the functions $F^{\pm}(\delta,h)$ occurring in each exponential are given by the analytical formula:

$$F^{\pm}(\delta,h)=\frac{(1\pm h)^2-\beta}{2+\delta\pm 2h}\pm\frac{\beta}{2\delta h}\left[1-\frac{(1-h^2)(1+\delta)}{(1-h^2+\delta)}\times\left(1\pm\frac{2h}{\sqrt{(1-h^2)(1+\delta^{-1})}}\arctan\sqrt{\frac{\sqrt{\delta^{-1}(1-h^2)+1}\pm h}{\sqrt{\delta^{-1}(1-h^2)+1}\mp h}}\right)^{-1}\right] \tag{19}$$

and $\beta\approx 0.38$ is a numerical parameter (see Eq. (A.6), etc. in Appendix A). For zero STT, $J$ = 0, Eq. (18) reduces to the known results of the Néel-Brown theory [32,43] for classical magnetic moments with superimposed easy-plane and in-plane easy-axis anisotropies plus the Zeeman term due to the applied magnetic field. In contrast to zero STT, for normalized spin currents $J\neq 0$, $\tau$ depends on $\alpha$ not only through the depopulation factors $A(\alpha S_{E_C}^{\pm})$ but also through the spin-polarized current *dependent* effective barrier heights $\Delta E^{\pm}$. This is so because parts of the arguments of the exponentials in Eq. (18) , namely Eq. (19), are markedly dependent on the ratio $J/\alpha$ and the dc bias field parameter. The turnover Eq. (18) also yields an asymptotic estimate for the inverse of the smallest nonvanishing eigenvalue of the Fokker-Planck operator $L_{FP}$ in Eq. (10). In addition, one may estimate two *individual* reversal times, namely, $\tau_{+}$ from the deeper well around the energy minimum at $\varphi_A=0$ and $\tau_{-}$ from the shallow well around the energy minimum at $\varphi_A=\pi$ (see Fig. 1(b)) as

$$\tau_{\pm}\approx\frac{2\pi\tau_0 e^{\sigma(1\pm h)^2\mp\frac{J}{\alpha}F^{\pm}(\delta,h)}}{A(\alpha S_{E_C}^{\pm})\sqrt{(1\pm h+\delta)(1\pm h)}}. \tag{20}$$

The individual times are in general unequal, i.e., $\tau_{+}\neq\tau_{-}$. In deriving Eqs. (18) and (20), all terms of order $\alpha^2,\alpha J,J^2$, etc. are neglected. This hypothesis is true only for the *underdamped regime*, $\alpha$ < 1, and *weak* spin-polarized currents, $J$<<1. (Despite these restrictions as we will see below Eqs. (18) and (20) still yield accurate estimates for $\tau$ for much higher values of $J$). Now, $\tau$ can also be calculated *numerically* via the method of statistical moments developed in Ref. [26] whereby the solution of the Fokker-Planck equation (10) in configuration space is reduced to the task of solving an infinite hierarchy of differential-recurrence equations for the averaged spherical harmonics $\langle Y_{lm}(\vartheta,\varphi)\rangle(t)$ governing the magnetization relaxation. (The $Y_{lm}(\vartheta,\varphi)$ are the spherical harmonics [46], and the angular brackets denote the statistical averaging). Thus one can evaluate $\tau$ numerically via $\lambda_1$ of the Fokker-Planck operator $L_{FP}$ in Eq. (10) by using matrix continued fractions as described in Ref. [47]. We remark that the ranges of applicability of the escape rate theory and the matrix continued-fraction method are in a sense complementary because escape rate theory cannot be used for low potential barriers, $\Delta E<3$, while the matrix continued-fraction method encounters substantial computational difficulties for *very high* potential barriers $\Delta E>25$ in the VLD range, $\alpha<10^{-4}$. Thus, in the foregoing sense, numerical methods and escape rate theory are very useful for the determination of $\tau$ for low and very high potential barriers, respectively. Nevertheless, in certain (wide) ranges of model parameters both methods yield accurate results for the reversal time (here these ranges are $5<\sigma<30$, $\sigma\delta>3$, and $\alpha>10^{-4}$). Then the numerically exact benchmark solution provided by the matrix continued fraction method allows one to test the accuracy of the analytical escape rate equations given above.

## IV. Results and discussion

Throughout the calculations, the anisotropy and spin-polarization parameters will be taken as $D_{\|}=0.034$, $\delta=20$, and $P=0.3$ ($P\approx 0.3\div 0.4$ are typical of ferromagnetic metals) just as in Ref. 6. Thus for $\gamma=2.2\times 10^5\,\mathrm{mA^{-1}s^{-1}}$, $T=300$ K, $v\sim 10^{-24}$ m$^3$, and a current density of the order of $\sim 10^7\,\mathrm{A\,cm^{-2}}$ in a 3 nm thick layer of cobalt with $M_S\approx 1.4\times 10^6\,\mathrm{Am^{-1}}$, we have the following estimates for the anisotropy (or inverse temperature) parameter $\sigma\approx 20.2$, characteristic time $\tau_0=(2\gamma M_S D_{\|})^{-1}\approx 0.48$ ps, and spin-polarized current parameter $J=b_P\hbar I/(|e|kT)\sim 1$. In Figs. 2 and 3, we compare $\tau$ from the asymptotic escape rate Eq. (18) with $\lambda_1^{-1}$ of the Fokker–Planck operator as calculated numerically via matrix continued fractions [26]. Apparently, $\tau$ as rendered by the turnover equation (18) and $\lambda_1^{-1}$ both lie very close to each other in the high barrier limit, where the asymptotic Eq. (18) provides an accurate approximation to $\lambda_1^{-1}$. In Fig. 2, $\tau$ is plotted as a function of $\alpha$ for various $J$. As far as STT effects are concerned they are governed by the ratio $J/\alpha$ so that by altering $J/\alpha$ the ensuing variation of $\tau$ may exceed several orders of magnitude (Fig. 2). Invariably for $J$ << 1, which is a condition of applicability of the escape rate equations (1) and (18), STT effects on the magnetization relaxation are pronounced only at very low damping, $\alpha$ << 1. For $\alpha\geq 1$, i.e. high damping, STT influences the reversal process very weakly because the STT term in Eq. (5) is then *small compared to the damping and random field terms.* Furthermore, $\tau$ may greatly exceed or, on the other hand, be very much less than the value for zero STT, i.e., $J$ = 0 (see Fig. 2). For example, as $J$ decreases from positive values, $\tau$ exponentially increases attaining a maximum at a critical value of the spin-polarized current and then smoothly switches over to exponential decrease as $|J|$ is further increased through negative values of $J$ [26]. Now, the temperature, external d.c. bias field, and damping dependence of $\tau$ can readily be understood in terms of the *effective potential barriers* $\Delta E^{\pm}$ in Eq. (18). For example, for $\sigma > 5$, the *temperature dependence* of $\tau$ has the customary Arrhenius behavior $\tau \sim e^{\Delta E^{-}}$, where $\Delta E^{-}$, Eq. (19), is markedly dependent on $J/\alpha$ (see Fig. 3a). Furthermore, the slope of $\tau(T^{-1})$ significantly decreases as the dc bias field parameter $h$ increases due to lowering of the barrier height $\Delta E^{-}$ owing to the action of the external field (see Fig. 3b). Now, although the range of applicability of Eqs. (18) and (20) is ostensibly confined to weak spin-polarized currents, $J << 1$, they can still yield accurate estimates for the reversal time for much higher values of $J$ far exceeding this condition (see Fig. 3a).

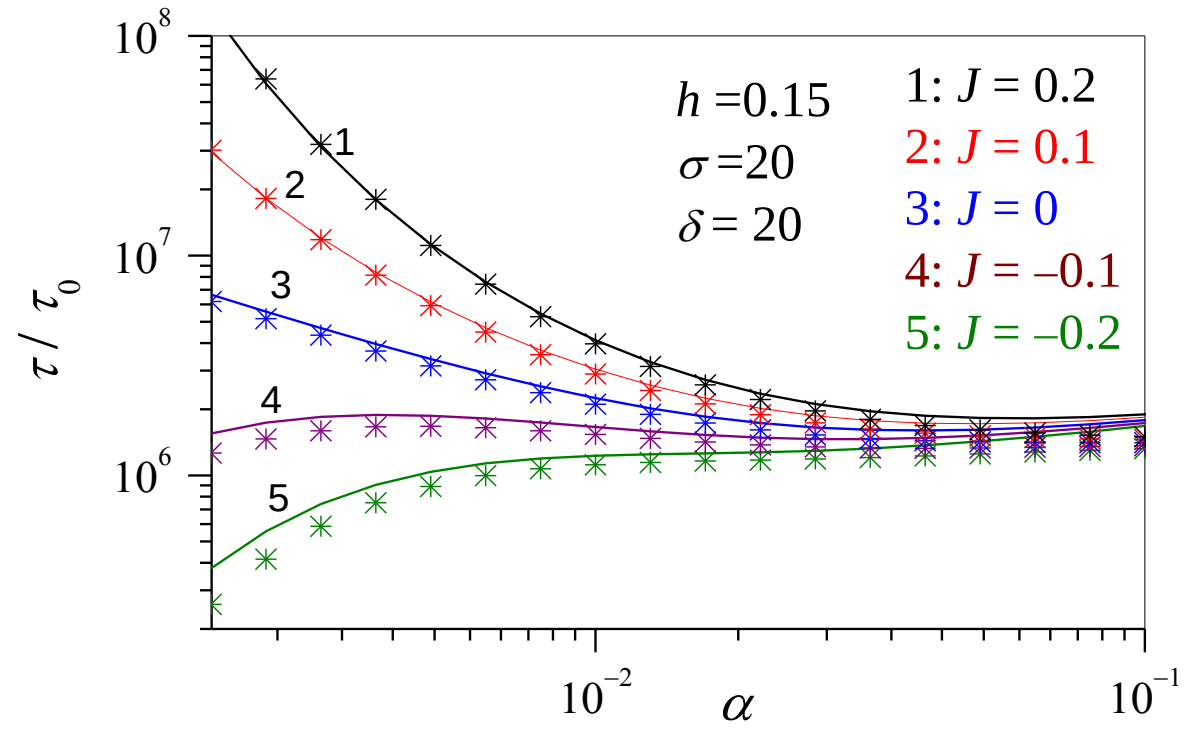

Fig. 2. Reversal time $\tau / \tau_0$ vs the damping parameter $\alpha$ for various values of the spin-polarized current parameter *J*. Solid lines: numerical calculations of the inverse of the smallest nonvanishing eigenvalue $(\tau_0\lambda_1)^{-1}$ of the Fokker–Planck operator, Eq. (11). Asterisks: the turnover formula, Eq. (18).

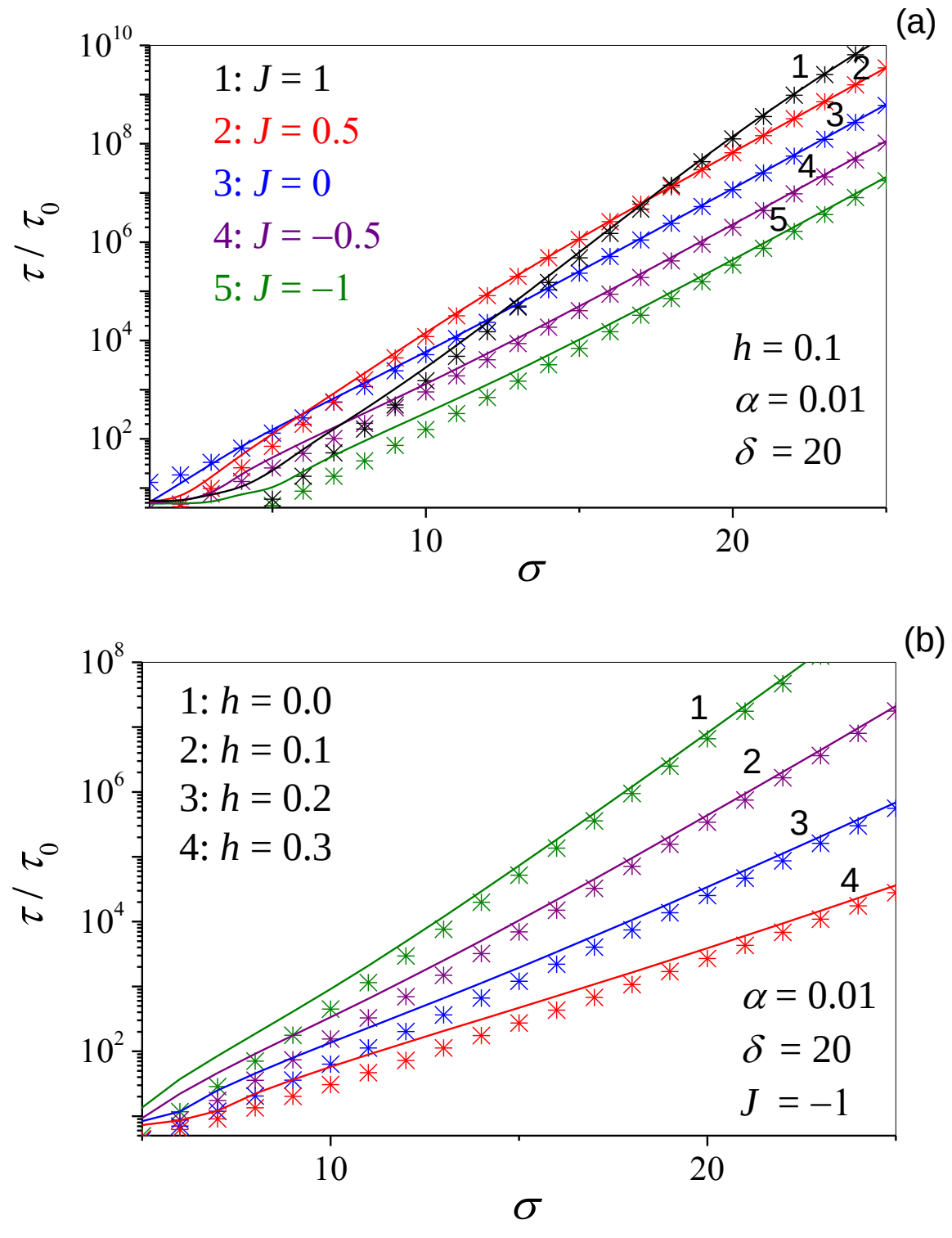

Fig. 3. Reversal time $\tau / \tau_0$ vs. the anisotropy (inverse temperature) parameter $\sigma$ for various spin-polarized currents *J* (a) and dc bias field parameters *h* (b). Solid lines: numerical solution for the inverse of the smallest nonvanishing eigenvalue $(\tau_0\lambda_1)^{-1}$ of the Fokker–Planck operator, Eq. (11). Asterisks: Eq. (18).

Thus, the turnover formula for $\tau$, Eqs. (18) and (20), bridging the Kramers VLD and ID escape rates as a function of the damping parameter for point particles [35,39-41] as extended by Coffey *et al.* [42,43] to the magnetization relaxation in nanoscale ferromagnets allows us (via the further extension to include STT embodied in Eq. (18)) to accurately evaluate STT effects in the magnetization reversal time of a nanomagnet driven by spin-polarized current in the highly relevant ID to VLD damping range. This (underdamped) range is characterized by $\alpha \le 1$ and the asymptotic escape rates are in complete agreement with independent numerical results [17]. Two particular merits of the escape rate equations for the reversal time are that (i) they are relatively simple (i.e., expressed via elementary functions) and (ii) that they can be used in those parameter ranges, where numerical methods (such as matrix continued fractions [17]) may be no longer applicable, e.g., for *very high barriers*, $\Delta E > 25$. Hence, one may conclude that the damping dependence of the magnetization reversal time is very marked in the underdamped regime $\alpha < 1$, a fact which may be very significant in interpreting many STT experiments.

## V. Appendix A: Calculation of $F^{\pm}(\delta, h)$ in Eq. (19)

For the bistable potential given by Eq. (7), and the nonconservative potential, Eq. (9), the spin-polarized current dependent effective barrier heights $\Delta E^{\pm}$ for each of the two wells are given by (cf. Eq. (12))

$$\Delta E^{\pm} = \sigma(1 \pm h)^2 \mp J\alpha^{-1}F^{\pm}(\delta,h), \tag{A.1}$$

where

$$F^{\pm}(\delta,h) = \sigma \int_{\varepsilon_A^{\pm}}^{\varepsilon_C} \frac{V_\varepsilon^{\pm}}{S_\varepsilon^{\pm}} d\varepsilon, \tag{A.2}$$

with $\varepsilon = E/\sigma$, $\varepsilon_A^{\pm} = E_A^{\pm}/\sigma = -1 \mp 2h$, $\varepsilon_C = E_C/\sigma = h^2$. The dimensionless action $S_\varepsilon^{+}$ and the dimensionless work done by the STT $V_\varepsilon^{+}$ for the deeper well can be calculated analytically via elliptic integrals as described in detail in Ref. [17] yielding

$$S_\varepsilon^{+} = \oint_\varepsilon \left( \left[ \frac{\partial E}{\partial \mathbf{u}} \times \mathbf{u} \right] \cdot d\mathbf{u} \right) = \frac{\sigma p_\varepsilon^2 (1+\delta)}{\tau_0 f_\varepsilon^{+}} \left\{ \varepsilon - 2hp_\varepsilon - \frac{h^2}{1+\delta} + \frac{1+\delta-h^2}{(1+\delta)(1+q_\varepsilon)} \left[ q_\varepsilon - 1 + \frac{2E(m_\varepsilon)}{(q_\varepsilon + m_\varepsilon)K(m_\varepsilon)} \right] + \left[ 4hp_\varepsilon - 2\frac{(1+\delta-h^2)(q_\varepsilon^2 - m_\varepsilon)}{(1+\delta)(1+q_\varepsilon)(q_\varepsilon+m_\varepsilon)} \right] \frac{\Pi(-q_\varepsilon | m_\varepsilon)}{K(m_\varepsilon)} \right\}, \tag{A.3}$$

$$V_\varepsilon^+ = \oint_\varepsilon \left( \left[ \mathbf{e}_p \times \mathbf{u} \right] \cdot d\mathbf{u} \right) = \frac{1}{2\tau_0 f_\varepsilon^+} \left\{ h + \frac{h^3}{(\delta+1)^2} - \frac{h(\varepsilon+1)}{\delta+1} \right.$$
$$+ p_\varepsilon \left( \varepsilon + 1 - \frac{2h^2}{\delta+1} \right) \left[ 2\frac{\Pi(-q_\varepsilon \mid m_\varepsilon)}{K(m_\varepsilon)} - 1 \right] \quad \text{(A.4)}$$
$$\left. + \frac{hp_\varepsilon^2}{1+q_\varepsilon} \left[ q_\varepsilon - 1 + 2\frac{q_\varepsilon E(m_\varepsilon) - (q_\varepsilon^2 - m_\varepsilon)\Pi(-q_\varepsilon \mid m_\varepsilon)}{(q_\varepsilon + m_\varepsilon)K(m_\varepsilon)} \right] \right\},$$

where

$$p_\varepsilon^2 = \frac{\delta - \varepsilon}{\delta+1} + \frac{h^2}{(\delta+1)^2}, \; q_\varepsilon = \frac{1-e_+}{1+e_+},$$

$$m_\varepsilon = \frac{(1+e_-)(1-e_+)}{(1+e_+)(1-e_-)}, \; e_\pm = -\frac{h\delta}{p_\varepsilon(\delta+1)} \pm \frac{\sqrt{h^2-\varepsilon}}{p_\varepsilon},$$

$K(m)$, $E(m)$, and $\Pi(a \mid m)$ are the *complete* elliptic integrals of the first, second, and third kinds, respectively [48], and $f_\varepsilon^+$ is the precession frequency in the deeper well at a given energy, namely,

$$f_\varepsilon^+ = \frac{p_\varepsilon \sqrt{(\delta+1)(1+e_+)(1-e_-)}}{8\tau_0 K(m_\varepsilon)}. \quad \text{(A.5)}$$

The quantities $S_\varepsilon^-$, $V_\varepsilon^-$, and $f_\varepsilon^-$ for the shallower well are obtained simply by replacing the dc bias field parameter $h$ by $-h$ in all the equations for $S_\varepsilon^+$, $V_\varepsilon^+$, and $f_\varepsilon^+$. We remark that $S_\varepsilon^\pm$ and $V_\varepsilon^\pm$ in Eqs. (A.3) and (A.4) differ by a factor 2 from those given in Ref. [17]. This is so because $S_\varepsilon^\pm$ and $V_\varepsilon^\pm$ are now calculated *between the saddle points and not over the precession period*. When $\varepsilon(\vartheta,\varphi) = \varepsilon_C$, $S_\varepsilon^\pm$ in Eqs. (A.3) reduces to $S_{E_C}^\pm$, Eq. (15).

In the parameter ranges $0 < h < 1$ and $\delta > 1$, the integral in Eq. (A.2) can be accurately evaluated analytically using an interpolation function for $V_\varepsilon^\pm / S_\varepsilon^\pm$ between the two limiting values $V_{\varepsilon_A}^\pm / S_{\varepsilon_A}^\pm$ and $V_{\varepsilon_C}^\pm / S_{\varepsilon_C}^\pm$ at $\varepsilon_A^\pm = -1 \mp h$ and $\varepsilon_C = h^2$, namely

$$\frac{V_\varepsilon^\pm}{S_\varepsilon^\pm} \approx \frac{V_{\varepsilon_A}^\pm}{S_{\varepsilon_A}^\pm} + \left( \frac{V_{\varepsilon_C}^\pm}{S_{\varepsilon_C}^\pm} - \frac{V_{\varepsilon_A}^\pm}{S_{\varepsilon_A}^\pm} \right) \left( \frac{\varepsilon - \varepsilon_A}{\varepsilon_C - \varepsilon_A} \right)^{\beta^{-1}-1}, \quad \text{(A.6)}$$

where $\beta \approx 0.38$ is an interpolation parameter yielding the best fit of $V_\varepsilon^\pm / S_\varepsilon^\pm$ in the interval $\varepsilon_A \le \varepsilon \le \varepsilon_C$. These limiting values can be calculated from Eqs. (A.3) and (A.4) yielding after tedious algebra:

$$\sigma \frac{V_{\varepsilon_A}^\pm}{S_{\varepsilon_A}^\pm} = \frac{1}{2+\delta \pm 2h} \quad \text{(A.7)}$$

and

$$\sigma \frac{V_{\varepsilon_C}^\pm}{S_{\varepsilon_C}^\pm} = \pm \frac{1}{2\delta h} \left\{ 1 - \frac{(1-h^2)(1+\delta)}{(1-h^2+\delta)} \right.$$
$$\left. \times \left[ 1 \pm \frac{2h}{\sqrt{(1-h^2)(1+\delta^{-1})}} \arctan \sqrt{\frac{\sqrt{\delta^{-1}(1-h^2)+1} \pm h}{\sqrt{\delta^{-1}(1-h^2)+1} \mp h}} \right]^{-1} \right\}. \quad \text{(A.8)}$$

Hence with Eqs. (A.2) and (A.6), we have a simple analytic formula for the current-dependent parts of the exponentials in Eq. (18) $F^\pm(\delta,h)$, viz.

$$F^\pm(\delta,h) \approx \left[ (1\pm h)^2 - \beta \right] \sigma \frac{V_{\varepsilon_A}^\pm}{S_{\varepsilon_A}^\pm} + \beta\sigma \frac{V_{\varepsilon_C}^\pm}{S_{\varepsilon_C}^\pm}, \quad \text{(A.9)}$$

which yields Eq. (19). For zero dc bias field, $h = 0$, Eq. (A.9) becomes

$$F^-(\delta,0) = F^+(\delta,0) \approx \frac{1-\beta}{2+\delta} + \frac{\beta\pi}{4\sqrt{\delta(\delta+1)}}. \quad \text{(A.10)}$$

The maximum relative deviation between the exact Eqs. (A.2) and approximate Eqs. (A.9) and (A.10) is less than 5% in the worst cases.

## References

[1] J. C. Slonczewski, “Current-driven excitation of magnetic multilayers”, *J. Magn. Magn. Mater*., vol. 159, p. L1, 1996.

[2] L. Berger, “Emission of spin waves by a magnetic multilayer traversed by a current”, *Phys. Rev. B*, vol. 54, p. 9353, 1996.

[3] M. D. Stiles and J. Miltat, “Spin-Transfer Torque and Dynamics”, in: *Spin Dynamics in Confined Magnetic Structures III*, p. 225, Eds. B. Hillebrandsn and A. Thiaville, Springer-Verlag, Berlin, 2006.

[4] D. C. Ralph and M. D. Stiles, Spin transfer torques, *J. Magn. Magn. Mater*. vol. 320, p. 1190, 2008.

[5] Y. Suzuki, A. A. Tulapurkar, and C. Chappert, “Spin-Injection Phenomena and Applications”, Ed. T. Shinjo, *Nanomagnetism and Spintronics.* Elsevier, Amsterdam, 2009, Chap. 3, p. 94.

[6] G. Bertotti, C. Serpico, and I. D. Mayergoyz, *Nonlinear Magnetization Dynamics in Nanosystems.* Elsevier, Amsterdam, 2009.

[7] T. Dunn, A. L. Chudnovskiy, and A. Kamenev, “Dynamics of nano-magnetic oscillators”, in *Fluctuating Nonlinear Oscillators*, Ed. M. Dykman. Oxford University Press, London, 2012.

[8] Yu. V. Gulyaev, P. E. Zilberman, A. I. Panas, and E. M. Epshtein, Spintronics: exchange switching of ferromagnetic metallic junctions at a low current density, *Usp. Fiz. Nauk*, vol. 179, p. 359, 2009 [*Phys. Usp*., vol. 52, p. 335, 2009].

[9] R. Heindl, W. H. Rippard, S. E. Russek, M. R. Pufall, and A. B. Kos, “Validity of the thermal activation model for spin-transfer torque switching in magnetic tunnel junctions”, *J. Appl. Phys*., vol. 109, p. 073910 (2011); W. H. Rippard, R. Heindl, M.R. Pufall, S. E. Russek, and A. B. Kos, ‘Thermal relaxation rates of magnetic nanoparticles in the presence of magnetic fields and spin-transfer effects’, *Phys. Rev. B*, vol. 84, p. 064439, 2011.

[10] J. Swiebodzinski, A. Chudnovskiy, T. Dunn, and A. Kamenev, “Spin torque dynamics with noise in magnetic nanosystems”, *Phys. Rev. B*, vol. 82, p. 144404, 2010.

[11] (a) E. B. Myers, F. J. Albert, J. C. Sankey, E. Bonet, R. A. Buhrman, and D. C. Ralph, “Thermally Activated Magnetic Reversal Induced by a Spin-Polarized Current”, *Phys. Rev. Lett*., vol. 89, p. 196801, 2002; (b) A. Fabian, C. Terrier, S. Serrano Guisan, X. Hoffer, M. Dubey, L. Gravier, J.-Ph. Ansermet, and J.-E. Wegrowe, “Current-Induced Two-Level Fluctuations in Pseudo-Spin-Valve (Co/Cu/Co) Nanostructures”, *Phys. Rev. Lett*., vol. 91, p. 257209, 2003; (c) F. B. Mancoff, R. W. Dave, N. D. Rizzo, T. C. Eschrich, B. N. Engel, and S. Tehrani, “Angular dependence of spin-transfer switching in a magnetic nanostructure”, *Appl. Phys. Lett*., vol. 83, p. 1596, 2003; (d) I. N. Krivorotov, N. C. Emley, A. G. F. Garcia, J. C. Sankey, S. I. Kiselev, D. C. Ralph, and R. A. Buhrman, “Temperature Dependence of Spin-Transfer-Induced Switching of Nanomagnets”, *Phys. Rev. Lett*., vol. 93, p. 166603, 2004; (e) R. H. Koch, J. A. Katine, and J. Z. Sun, “Time-Resolved Reversal of Spin-Transfer Switching in a Nanomagnet”, *Phys. Rev. Lett*., vol. 92, p. 088302, 2004; (f) M. L. Schneider, M. R. Pufall, W. H. Rippard, S. E. Russek, and J. A. Katine, “Thermal effects on the critical current of spin torque switching in spin valve nanopillars”, *Appl. Phys. Lett*., vol. 90, p. 092504, 2007.

[12] W. T. Coffey and Y. P. Kalmykov, *The Langevin Equation*, 4th ed. World Scientific, Singapore, 2017.

[13] Z. Li and S. Zhang, “Thermally assisted magnetization reversal in the presence of a spin-transfer torque”, *Phys. Rev. B*, vol. 69, p. 134416, 2004.

[14] D. M. Apalkov and P. B. Visscher, “Spin-torque switching: Fokker-Planck rate calculation”, *Phys. Rev. B*, vol. 72, p. 180405(R), 2005.

[15] G. Bertotti, I. D. Mayergoyz, and C. Serpico, "Analysis of random Landau-Lifshitz dynamics by using stochastic processes on graphs", *J. Appl. Phys*., vol. 99, p. 08F301, 2006.

[16] T. Taniguchi and H. Imamura, "Thermally assisted spin transfer torque switching in synthetic free layers", *Phys. Rev. B*, vol. 83, p. 054432, 2011; "Thermal switching rate of a ferromagnetic material with uniaxial anisotropy", *Phys. Rev. B*, vol. 85, p. 184403, 2012; T. Taniguchi, Y. Utsumi, and H. Imamura, "Thermally activated switching rate of a nanomagnet in the presence of spin torque", *Phys. Rev. B*, vol. 88, p. 214414, 2013; T. Taniguchi, Y. Utsumi, M. Marthaler, D. S. Golubev, and H. Imamura, "Spin torque switching of an in-plane magnetized system in a thermally activated region", *Phys. Rev. B*, vol. 87, p. 054406, 2013; T. Taniguchi and H. Imamura, "Current dependence of spin torque switching rate based on Fokker-Planck approach", *J. Appl. Phys*., vol. 115, p. 17C708, 2014.

[17] D. Byrne, W.T. Coffey, W. J. Dowling, Y.P. Kalmykov, and S.T. Titov, "Spin transfer torque and dc bias magnetic field effects on the magnetization reversal time of nanoscale ferromagnets at very low damping: Mean first-passage time versus numerical methods", *Phys. Rev. B*, vol. 93, p. 064413, 2016.

[18] W.H. Butler, T. Mewes, C.K.A. Mewes, P.B. Visscher, W.H. Rippard, S.E. Russek, and R. Heindl, "Switching Distributions for Perpendicular Spin-Torque Devices Within the Macrospin Approximation", *IEEE Trans. Magn.*, vol. 48, p. 4684, 2012.

[19] K. A. Newhall and E. Vanden-Eijnden, "Averaged equation for energy diffusion on a graph reveals bifurcation diagram and thermally assisted reversal times in spin-torque driven nanomagnets". *J. Appl. Phys*., vol. 113, p. 184105, 2013.

[20] J. M. Lee and S. H. Lim, "Thermally activated magnetization switching in a nanostructured synthetic ferrimagnet", *J. Appl. Phys.*, vol. 113, p. 063914, 2013.

[21] D. V. Berkov and N. L. Gorn, "Magnetization precession due to a spin-polarized current in a thin nanoelement: Numerical simulation study", *Phys. Rev. B*, vol. 72, p. 094401, 2005; "Non-linear magnetization dynamics in nanodevices induced by a spin-polarized current: micromagnetic simulation", *J. Phys. D: Appl. Phys*., vol. 41, p. 164013, 2008.

[22] K. Ito, "Micromagnetic Simulation on Dynamics of Spin Transfer Torque Magnetization Reversal", *IEEE Trans. Magn.*, vol. 41, p. 2630, 2005.

[23] T. Taniguchi, M. Shibata, M. Marthaler, Y. Utsumi and H. Imamura, "Numerical Study on Spin Torque Switching in Thermally Activated Region", *Appl. Phys. Express*, vol. 5, p. 063009, 2012.

[24] D. Pinna, A. D. Kent, and D. L. Stein, "Spin-Transfer Torque Magnetization Reversal in Uniaxial Nanomagnets with Thermal Noise", *J. Appl. Phys*., vol. 114, p. 033901, 2013; "Thermally assisted spin-transfer torque dynamics in energy space", *Phys. Rev. B*, vol. 88, p. 104405, 2013; "Spin-torque oscillators with thermal noise: A constant energy orbit approach", *Phys. Rev. B*, vol. 90, 174405, 2014; "Large fluctuations and singular behavior of nonequilibrium systems", *Phys. Rev. B*, vol. 93, p. 012114, 2016.

[25] M. d'Aquino, C. Serpico, R. Bonin, G. Bertotti, and I. D. Mayergoyz, "Stochastic resonance in noise-induced transitions between self-oscillations and equilibria in spin-valve nanomagnets", *Phys. Rev. B*, vol. 84, p. 214415, 2011.

[26] Y. P. Kalmykov, W. T. Coffey, S. V. Titov, J. E. Wegrowe, and D. Byrne, "Spin-torque effects in thermally assisted magnetization reversal: Method of statistical moments", *Phys. Rev. B*, vol. 88, p. 144406, 2013; D. Byrne, W. T. Coffey, Y. P. Kalmykov, S. V. Titov, and J. E. Wegrowe, "Spin-transfer torque effects in the dynamic forced response of the magnetization of nanoscale ferromagnets in superimposed ac and dc bias fields in the presence of thermal agitation", *Phys. Rev. B*, vol. 91, p. 174406, 2015.

[27] L. Néel, "Théorie du traînage magnétique des ferromagnétiques en grains fins avec applications aux terres cuites », *Ann. Géophys*., vol. 5, p. 99, 1949.

[28] W. F. Brown, Jr., "Thermal fluctuations of a single-domain particle", *Phys. Rev*., vol. 130, p. 1677, 1963.

[29] W.F. Brown Jr., "Thermal Fluctuations of Fine Ferromagnetic Particles", *IEEE Trans. Mag.*, vol. 15, p. 1196, 1979.

[30] I. Klik and L. Gunther, "First-passage-time approach to overbarrier relaxation of magnetization", *J. Stat. Phys*., vol. 60, p. 473, 1990; I. Klik and L. Gunther, "Thermal relaxation over a barrier in single domain ferromagnetic particles", *J. Appl. Phys*., vol. 67, p. 4505, 1990.

[31] W. T. Coffey, "Finite integral representation of characteristic times of orientation relaxation processes: Application to the uniform bias force effect in relaxation in bistable potentials", *Adv. Chem. Phys*., vol. 103, p. 259, 1998.

[32] W. T. Coffey and Y. P. Kalmykov, "Thermal fluctuations of magnetic nanoparticles", *J. Appl. Phys*., vol. 112, p. 121301, 2012.

[33] H. A. Kramers, "Brownian motion in a field of force and the diffusion model of chemical reactions", *Physica*, vol. 7, p. 284, 1940.

[34] P. Hänggi, P. Talkner, and M. Borkovec, "Reaction-Rate Theory: Fifty Years After Kramers", *Rev. Mod. Phys*., vol. 62, p. 251, 1990.

[35] W.T. Coffey, Y.P. Kalmykov, and S.T. Titov, "Magnetization reversal time of magnetic nanoparticles at very low damping", *Phys. Rev. B*, vol. 89, p. 054408, 2014; D. Byrne, W.T. Coffey, W. J. Dowling, Y.P. Kalmykov, and S.T. Titov, "On the Kramers very low damping escape rate for point particles and classical spins", *Adv. Chem. Phys*., vol. 156, p. 393, 2015.

[36] W. Wernsdorfer, E. Bonet Orozco, K. Hasselbach, A. Benoit, D. Mailly, O. Kubo, H. Nakano, and B. Barbara, "Macroscopic Quantum Tunneling of Magnetization of Single Ferrimagnetic Nanoparticles of Barium Ferrite", *Phys. Rev. Lett*., vol. 79, p. 4014, 1997; W.T. Coffey, D.S.F. Crothers, J.L. Dormann, Yu.P. Kalmykov, E.C. Kennedy, and W. Wernsdorfer, "Thermally Activated Relaxation Time of a Single Domain Ferromagnetic Particle Subjected to a Uniform Field at an Oblique Angle to the Easy Axis: Comparison with Experimental Observations", *Phys. Rev. Lett*., vol. 80, p. 5655, 1998.

[37] M. Oogane, T. Wakitani, S. Yakata, R. Yilgin, Y. Ando, A. Sakuma, and T. Miyazaki, "Magnetic Damping in Ferromagnetic Thin Films", *Jpn. J. Appl. Phys*., vol. 45, p. 3889, 2006.

[38] M. C. Hickey and J. S. Moodera, "Origin of Intrinsic Gilbert Damping", *Phys. Rev. Lett*., vol. 102, p. 137601 (2009).

[39] V. I. Mel'nikov and S. V. Meshkov, "Theory of activated rate processes: exact solution of the Kramers problem", *J. Chem. Phys*., vol. 85, p. 1018, 1986.

[40] H. Grabert, "Escape from a metastable well: The Kramers turnover problem", *Phys. Rev. Lett.*, vol. 61, p. 1683, 1988.

[41] E. Pollak, H. Grabert, and P. Hänggi, "Theory of activated rate processes for arbitrary frequency dependent friction: Solution of the turnover problem", *J. Chem. Phys*., vol. 91, p. 4073, 1989.

[42] W. T. Coffey, D. A. Garanin, and D. J. McCarthy, "Crossover formulas in the Kramers theory of thermally activated escape rates – application to spin systems", *Adv. Chem. Phys*. vol. 117, p. 483, 2001; P.M. Déjardin, D.S.F. Crothers, W.T. Coffey, and D.J. McCarthy, "Interpolation formula between very low and intermediate-to-high damping Kramers escape rates for single-domain ferromagnetic particles", *Phys. Rev. E*, vol. 63, p. 021102, 2001.

[43] Yu. P. Kalmykov, W.T. Coffey, B. Ouari, and S. V. Titov, "Damping dependence of the magnetization relaxation time of single-domain ferromagnetic particles", *J. Magn. Magn. Mater*., vol. 292, p. 372, 2005; Yu. P. Kalmykov and B. Ouari, "Longitudinal complex magnetic susceptibility and relaxation times of superparamagnetic particles with triaxial anisotropy", *Phys. Rev. B*, vol. 71, p. 094410, 2005; B. Ouari and Yu. P. Kalmykov, "Dynamics of the magnetization of single domain particles having triaxial anisotropy subjected to a uniform dc magnetic field", *J. Appl. Phys*., vol. 100, p. 123912, 2006.

[44] T. L. Gilbert, "A Lagrangian formulation of the gyromagnetic equation of the magnetic field", *Phys. Rev*., vol. 100, p. 1243, 1955 (Abstract only; full report in: Armour Research Foundation Project No. A059, Supplementary Report, 1956). Reprinted in T.L. Gilbert, A phenomenological theory of damping in ferromagnetic materials, *IEEE Trans. Magn*., vol. 40, p. 3443, 2004.

[45] J. Z. Sun, "Spin-current interaction with a monodomain magnetic body: A model study", *Phys. Rev. B*, vol. 62, p. 570, 2000.

[46] D. A. Varshalovitch, A. N. Moskalev, and V. K. Khersonskii, *Quantum Theory of Angular Momentum*. World Scientific, Singapore, 1988.

[47] Y.P. Kalmykov, "Evaluation of the smallest nonvanishing eigenvalue of the Fokker-Planck equation for the Brownian motion in a potential. II. The matrix continued fraction approach", *Phys. Rev. E*, vol. 62, p. 227, 2000.

[48] M. Abramowitz and I. Stegun, Eds., *Handbook of Mathematical Functions.* Dover, New York, 1964.

```
In[227]:=
AS[S_] := Exp[1/Pi * NIntegrate[Log[1. - Exp[-S * (r^2 + 0.25)]]/(r^2 + 0.25), {r, 0., 31.}, AccuracyGoal -> 8]];

ep[δ_, h_, ϵ_] := (-(h*δ)/(1+δ) + Sqrt[h^2 - ϵ])/Sqrt[(δ-ϵ)/(1+δ) + h^2/(1+δ)^2]; ap[δ_, h_, ϵ_] := (1 - ep[δ, h, ϵ])/(1 + ep[δ, h, ϵ]); me[δ_, h_, ϵ_] := (1 - ep[δ, -h, ϵ])/(1 + ep[δ, h, ϵ]) * (1 - ep[δ, h, ϵ])/(1 + ep[δ, -h, ϵ]);

Te[δ_, h_, ϵ_] := (8 * EllipticK[me[δ, h, ϵ]])/Sqrt[((δ - ϵ + h^2/(1+δ)) * (1 + ep[δ, h, ϵ]) * (1 + ep[δ, -h, ϵ]))];

Ve[δ_, h_, ϵ_] :=
  0.5 * Te[δ, h, ϵ] * (h + h^3/(δ + 1)^2 - h (ϵ + 1)/(δ + 1) + (ϵ + 1 - 2 h^2/(δ + 1)) Sqrt[(δ - ϵ)/(1 + δ) + h^2/(1 + δ)^2] (-1 + (2 EllipticPi[-ap[δ, h, ϵ], me[δ, h, ϵ]])/EllipticK[me[δ, h, ϵ]]) + (h * ((δ-ϵ)/(1+δ) + h^2/(1+δ)^2))/(ap[δ, h, ϵ] + 1) * (ap[δ, h, ϵ] - 1 +
    (2 (ap[δ, h, ϵ] EllipticE[me[δ, h, ϵ]] - (ap[δ, h, ϵ]^2 - me[δ, h, ϵ]) EllipticPi[-ap[δ, h, ϵ], me[δ, h, ϵ]]))/((ap[δ, h, ϵ] + me[δ, h, ϵ]) EllipticK[me[δ, h, ϵ]])));

Sp[δ_, h_, ϵ_] :=
  Te[δ, h, ϵ] * (δ - ϵ + h^2/(1 + δ)) * σ * (ϵ - h^2/(δ + 1) - 2 h Sqrt[(δ - ϵ)/(1 + δ) + h^2/(1 + δ)^2] (1 - (2 EllipticPi[-ap[δ, h, ϵ], me[δ, h, ϵ]])/EllipticK[me[δ, h, ϵ]]) + 1/(ap[δ, h, ϵ] + 1) (1 - h^2/(1 + δ)) (ap[δ, h, ϵ] - 1 +
    2 * (ap[δ, h, ϵ] EllipticE[me[δ, h, ϵ]] - (ap[δ, h, ϵ]^2 - me[δ, h, ϵ]) EllipticPi[-ap[δ, h, ϵ], me[δ, h, ϵ]])/((ap[δ, h, ϵ] + me[δ, h, ϵ]) EllipticK[me[δ, h, ϵ]])));


τturnover[σ_, δ_, h_, α_, J_] := Block[{NP, NM, s1, S2, Γ1, Γ2}, s1 = (4 * δ * (1 - h^2 + δ) * σ)/(1 + δ)^(3/2) * (Sqrt[((1 - h^2) * (1 + δ))/δ] + 2 * h * ArcTan[Sqrt[(h δ + Sqrt[δ (1 - h^2 + δ)])/(-h δ + Sqrt[δ (1 - h^2 + δ)])]]);
  s2 = (4 * δ * (1 - h^2 + δ) * σ)/(1 + δ)^(3/2) * (Sqrt[((1 - h^2) * (1 + δ))/δ] - 2 * h * ArcTan[Sqrt[(-h δ + Sqrt[δ (1 - h^2 + δ)])/(h δ + Sqrt[δ (1 - h^2 + δ)])]]); Np = NIntegrate[Ve[δ, h, ϵ]/Sp[δ, h, ϵ], {ϵ, -1 - 2 h, h^2}];
  Nm = NIntegrate[Ve[δ, -h, ϵ]/Sp[δ, -h, ϵ], {ϵ, -1 + 2 h, h^2}]; Γ1 = Sqrt[(1 + h + δ) (1 + h)]/(2 * π) * Exp[-σ * ((1 + h)^2 - J/α * Np)]; Γ2 = Sqrt[(1 - h + δ) (1 - h)]/(2 * π) * Exp[-σ * ((1 - h)^2 + J/α * Nm)];
  Return[0.5/(Γ1 + Γ2) * AS[α * s1 + α * s2]/(AS[α * s1] * AS[α * s2])];];
δ = 20.; h = 0.2; α = 0.01; σ = 20.; J = -0.1; X = h^2 - 0.000000001; β = 0.38;
s = (4 * δ * (1 - h^2 + δ) * σ)/(1 + δ)^(3/2) * (Sqrt[((1 - h^2) * (1 + δ))/δ] + 2 * h * ArcTan[Sqrt[(h δ + Sqrt[δ (1 - h^2 + δ)])/(-h δ + Sqrt[δ (1 - h^2 + δ)])]])

st = Sp[δ, h, X]

v = (2 * (1 - h^2 + δ))/(1 + δ)^(3/2) * (h Sqrt[δ (1 + δ) (1 - h^2)]/(1 - h^2 + δ) + 2 * ArcTan[Sqrt[(h δ + Sqrt[δ (1 - h^2 + δ)])/(-h δ + Sqrt[δ (1 - h^2 + δ)])]])

ve = Ve[δ, h, X]
test1 = ve/st

test2 = 1/(2 * δ * h * σ) * (1 - (((1-h^2) * (1+δ))/(1-h^2+δ))/(1 + 2 * h Sqrt[δ/((1-h^2) * (1+δ))] ArcTan[Sqrt[(h δ + Sqrt[δ (1-h^2+δ)])/(-h δ + Sqrt[δ (1-h^2+δ)])]]))

aaa = ((1 + h)^2 - β)/(2 + δ + h) + β/(2 * δ * h) * (1 - (((1-h^2) * (1+δ))/(1-h^2+δ))/(1 + 2 * h Sqrt[δ/((1-h^2) * (1+δ))] ArcTan[Sqrt[(h δ + Sqrt[δ (1-h^2+δ)])/(-h δ + Sqrt[δ (1-h^2+δ)])]]))

aaa = σ * NIntegrate[Ve[δ, h, ϵ]/Sp[δ, h, ϵ], {ϵ, -1 - 2 * h, h^2}]
TAU = τturnover[σ, δ, h, α, J]

Out[234]= 473.059
Out[235]= 473.059
Out[236]= 0.853362
Out[237]= 0.853362
Out[238]= 0.00180392
Out[239]= 0.00180392
Out[240]= 0.0614576
Out[241]= 0.060431
Out[242]= 320948.
```